\documentclass[article,useAMS,usenatbib]{mn2e}

\usepackage{mathtools}
\usepackage{graphicx}
\usepackage{float}

\title[Radio Imaging of $\gamma$-ray Nova V959 Mon]{Multi-Epoch Radio Imaging of $\gamma$-ray Nova V959 Mon}
\author[F. Healy, T. J. O'Brien, R. J. Beswick, A. Avison and M. K. Argo]{F. Healy$^{1}$, T. J. O' Brien$^{1}$, R. Beswick$^{1}$, A. Avison$^{1}$ and M. K. Argo$^{2,1}$\thanks{Email: fiona.healy@postgrad.manchester.ac.uk\newline
Tim.OBrien@manchester.ac.uk\newline
Robert.Beswick@manchester.ac.uk\newline
adam.avison@manchester.ac.uk\newline
margo@uclan.ac.uk}\\
$^{1}$Jodrell Bank Centre for Astrophysics, University of Manchester, Oxford Road, Manchester, M13 9PL, United Kingdom\\
$^{2}$Jeremiah Horrocks Institute, University of Central Lancashire,
Preston, Lancashire,
PR1 2HE,
United Kingdom}
\begin{document}

\date{Accepted 2017 May 3. Received 2017 May 2; in original form 2016 July 18}

\pagerange{\pageref{firstpage}--\pageref{lastpage}} \pubyear{2016}

\maketitle

\label{firstpage}

\begin{abstract}
V959 Mon (Nova Mon 2012) was first detected by the Fermi
Large Area Telescope in June 2012, as a transient gamma-ray 
source. Subsequent optical observations showed that this
gamma-ray emission was due to a classical nova explosion.
Multi-frequency observations of V959 Mon with the VLA
between June and September 2012 revealed dramatic brightening,
and a spectrum that steepened with increasing frequency.
High resolution radio images of V959 Mon using e-MERLIN
are presented here, at six epochs between September 2012
and February 2014 which show morphological evolution of
the source. While early e-MERLIN observations of V959 Mon
show an east-west elongation in the ejecta
morphology, subsequent observations suggest that the ejecta
become elongated in the north-south direction. Our
high-resolution observations of this surprising evolution
in the structure of V959 Mon can assist us in further
understanding the behaviour and morphology of nova ejecta.

\end{abstract}

\begin{keywords}
novae,cataclysmic variables - binaries - stars:individual(V959 Mon)
\end{keywords}

\section{Introduction}

\subsection{Classical Novae}

A nova is a cataclysmic variable star consisting of a
white dwarf (WD) and a main sequence, subgiant or red
giant companion. In a nova explosion, the white dwarf
undergoes a thermonuclear runaway (TNR) on its surface,
as a result of build-up of accreted material from its
companion. This leads to a large expulsion of matter
from the WD surface, as well as a dramatic increase in
the optical magnitude of the system. Novae typically expel
between $10^{-7}$ M$_{\odot}$ and $10^{-4}$ M$_{\odot}$
of matter during an explosion, at velocities that can
range from a few hundred to several thousand km s$^{-1}$ (see \citealp{Bode}).

Novae occur in the Milky Way at a rate of about 35 yr$^{-1}$
\citep{Shafter1}. Some novae have been observed to recur;
see, for example, \citet{Schaefer}. Such recurring novae
can be considered candidates for Type 1a supernovae, provided
the mass accreted from the companion between outbursts is
greater than the mass lost from the WD during outbursts.

Nova explosions emit across the electromagnetic spectrum.
In the optical, their light curves feature a sudden increase
in brightness (corresponding to the time of the explosion)
followed by a gradual decrease back to the pre-outburst
levels. Several classical novae have now been detected as
GeV $\gamma$-ray sources by the Fermi satellite, indicating
that relativistic particle acceleration (due to shocks)
takes place in nova ejecta (see \citealp{Fermi1}). Novae are
also radio sources, primarily due to thermal bremsstrahlung,
although some also show evidence of synchrotron radiation;
an indication of the presence of shocks in nova
explosions. Novae are slower to rise to maximum brightness
in the radio, with the peak magnitude occurring some time
(days to months) after the optical outburst.

Attempts have been made to establish a model of ejecta
expansion in novae, based on the observed radio light
curves. Most of these models, for example the
``Hubble flow'' model, \citep{Hjellming1} assume an
isothermal, smooth, spherical outflow. The Hubble flow
model specifically assumes that all the material is ejected
at once. It further assumes that different components of
the gas have different velocities. The result is a thick shell, with
the slowest parts of the gas at the
inner radius, and the fastest parts at the outer radius.
This model predicts a $t^2$ rise
in radio intensity, and a $t^{-3}$ decay at late times
once the source has become optically thin. In between
these times, the intensity decays as $t^{-\frac{4}{3}}$.
Simulated light curves using these assumptions have been
used to fit observed radio light curves of novae (see for
example \citealp{Seaquist1}). However, more recent radio
light curves of novae indicate rather poor agreement with
these simple models (see for example \citealp{Krauss}).
Furthermore, many novae which have been resolved and imaged
by radio instruments appear to display a non-spherical
morphology (e.g. \citealp{Taylor1}, \citealp{Eyres},
\citealp{Heywood1}). Some studies, for example
\citet{Ribeiro2} and \citet{Heywood2}, have used simulated radio emission to
investigate the effects of a non-spherical ejecta on the
nova light curve.

\subsection{V959 Mon}

V959 Mon, also known as Nova Mon 2012, was first detected
on June 19$^{\text{th}}$ 2012 by the Fermi Large Area
Telescope, as a GeV $\gamma$-ray transient source
\citep{Cheung2} named J0639+0548.
Twelve and sixteen days after this discovery, \citet{Chomiuk1}
obtained radio spectra of J0639+0548, using the Karl
G. Jansky Very Large Array (VLA). The spectra were
unusually flat for a nova -- novae usually emit in the
radio due to thermal bremsstrahlung (and at such an early
stage, the ejecta would be expected to be optically thick); however, a
flat spectrum such as the one observed is indicative of
synchrotron emission. Following the Fermi detection, V959 Mon
was obscured by the Sun, and thus could not be observed
optically (\citealp{Munari1}).

In August 2012, \citet{Fujikawa} detected an optical
nova in the same region of the sky as J0639+0548, with
a magnitude of V $\sim$ 9.4. This newly-discovered
nova was connected to the $\gamma$-ray source by
\citet{Cheung1}, who deduced that the $\gamma$-ray
emission discovered in J0639+0548 had been emitted
from the new nova, and that the GeV emission was likely to
have taken place around the beginning of the nova explosion.

From the time of V959 Mon's discovery as a nova, it has
been observed across many frequency bands, including X-ray,
UV, optical, IR and radio. Spectroscopic studies indicate
that V959 Mon is an ONe nova (see \citealp{Munari};
\citealp{Shore}). Swift XRT observations detected a
strong hard X-ray source with a temperature of
$4 \times 10^7$~K \citep{Nelson}. Both optical and
X-ray observations have confirmed a 7.1 hour orbital
period in V959 Mon (see \citealp{Osborne}, \citealp{Page},
\citealp{Wagner}). Through the comparison of synthetic
line spectra with line spectra observed in V959 Mon,
\citet{Ribeiro} suggested that V959 Mon had a bipolar
ejecta structure, with an angle of inclination
(defined as the angle between the orbital plane and
the plane of the sky) of $82^{\circ} \pm 6^{\circ}$. A high
inclination angle was also found by \citet{Page}.
The distance to V959 Mon was estimated by
\citet{Munari} to be 1.5 kpc. An estimate of the distance
was also made by \citet{Linford}, who found that the
distance lay between 0.9 $\pm$ 0.2 and 2.2 $\pm$ 0.4 kpc,
with a probable distance of 1.4 $\pm$ 0.4 kpc.

European Very Long Baseline Interferometry (VLBI)
Network (EVN) observations (made in September 2012) at
2-7 milliarcsecond (mas) resolution of V959 Mon at 5 GHz
by \citet{Tim1} found it to be a double radio source,
with 2 compact components extending to the
north-west and south-east. Observations from October
2012 showed that the two components had moved further
apart, with a proper motion of 0.45 milliarcseconds per
day. Subsequent observations (113 days after the Fermi
discovery of V959 Mon) detected a third VLBI component.
The brightest VLBI component was also resolved by the Very Long Baseline Array (VLBA).
It was found to have a peak brightness temperature of
$2 \times 10^{6}$K, indicative of non-thermal
(synchrotron) emission (see \citealp{Chomiuk1}).

Following the initial observations made by \citet{Chomiuk1},
VLA observations of V959 Mon continued up to about 615 days after the Fermi discovery.
126 days after the Fermi detection (day 0), a clearly
bipolar structure extending from east to west was observed.
However, 16 months later (day 615), the geometry of the
ejecta had changed from east-west to north-south. This
evolution was also described by \citet{Linford}, who describe VLA observations
between 126 and 199 days after the Fermi discovery, and 
between 615 and 703 days. \citet{Linford} also found
that between day 199 and day 615, the morphology of V959 Mon had changed
its orientation from an east-west elongation to a north-south elongation.
Light curves made using the observations of \citet{Chomiuk1}
were fit with model curves based on the ``Hubble flow'' model
\citep{Hjellming1}. Assuming an outer shell velocity of $2400~$km s$^{-1}$, it
was found that the models were best fit using values of
$200~$km s$^{-1}$ for the velocity of the inner radius of
the shell, $4 \times 10^{-5} M_{\odot}$ for the ejecta mass,
and an electron temperature of $2 \times 10^4$~K. While
this estimate of the ejecta mass is plausible, it was
noted by \citet{Chomiuk1} that the observed morphology
of V959 Mon was aspherical, and therefore that the
Hubble flow model was unlikely to provide a complete
description of V959 Mon's behaviour.

V959 Mon exhibits several traits that are contrary to the
standard models of mass ejection from novae. Most notable
are its emission of GeV $\gamma$-rays, and its aspherical
and variable ejecta -- first appearing as two components
extended from south-west to north-east, then as a bipolar
structure extending from east to west, and subsequently
undergoing a 90$^{\circ}$ shift, appearing as a bipolar
structure extending from north to south.

\citet{Chomiuk1} have proposed a theoretical explanation
for these features. Firstly, that the orbiting binary
system transfers some of its orbital energy to the nova
ejecta. They suggest that this leads to a slow expansion
of ejecta along the orbital plane (suggested to be north-south in the
case of V959 Mon and given its high inclination). At early times, this is difficult to
detect and resolve, as it expands very slowly. Secondly,
the white dwarf expels fast winds, which propagate more
easily along the low-density regions at the poles. This
fast-expanding body of ejecta, which would appear from east to west in the case of V959 Mon (perpendicular to the orbital
plane) would dominate radio images while optically thick.
Eventually this rapidly expanding ejecta
would become optically thin, leaving the slower-moving
north-south ejecta, still optically thick, to dominate radio images. This
model would explain V959 Mon's shift from an east-west
orientation to a north-south orientation.

The model proposed by \citet{Chomiuk1} would also provide
an explanation for the presence of $\gamma$ ray emission
and for the non-thermal VLBI components detected;
interaction between the slow-moving and fast-moving phases
of ejection would lead to strong shocks at their interface,
generating the high energies required to produce these two
effects.

\section{Observations}

The observations presented here were made with the
e-MERLIN array over 6 epochs: 2012 September 13 and 18,
2012 November 12, 14 and 15, 2012 November 22-24, 2013 February
26, 2013 October 11-14 and 2014 February 21-22.
All seven antennas in the array were used for the
February 2013 observations;
the Lovell telescope was omitted for the other epochs. The observations were made in the C
Band (4-8 GHz), providing a resolution of around 40
milliarcseconds, and with a bandwidth of 512 MHz. The
first three epochs had a central frequency of 5.7 GHz, and
the subsequent three had a central frequency of 5.0 GHz.
The bandwidth was divided into 4 intermediate frequency
bands (IFs), each containing 512 channels. The data
were averaged such that each IF effectively contained
128 channels. The total times spent observing the target source,
as well as the average scan lengths, are given in Table \ref{times}.
\begin{center}
\begin{table*}

\begin{tabular}{| c | c | c |}
\hline
Epoch &  \begin{tabular}{@{}c@{}}Time \\ on Source (hr)\end{tabular} &  \begin{tabular}{@{}c@{}}RMS noise level\\ in image (Jy/Beam)\end{tabular} \\
\hline
2012 Sep 13 and 18 & 8 & 7.029e-5\\
2012 Nov 12, 14 and 15 & 18.5 & 1.3e-4\\
2012 Nov 22-24 & 8.5 & 1.36e-4\\
2013 Feb 26 & 4 & 9.924e-5\\
2013 Oct 11-14 & 7 & 8.82e-5\\
2014 Feb 21-22 & 12 & 2.526e-5\\
\hline
\end{tabular}
\caption{\textit{Epochs at which the e-MERLIN observations of V959 Mon were made, total hours spent observing the target source
at each epoch, and image RMS noise levels (in Jy/Beam).}}
\label{times}

\end{table*}
\end{center}

The calibration, reduction and imaging of each epoch was
carried out with the Astronomical Image Processing System (AIPS), using standard tasks. In all cases,
J0645+0541 was observed as a phase calibrator, 3C286 as
a flux calibrator and OQ208 as a band-pass calibrator.
The images were constructed using the different
CLEAN restoring beams fitted to each one. They were then
re-constructed using the same 70 mas $\times$ 70 mas
circular beam for each epoch, so that the morphology
could be more reliably compared from epoch to epoch.

\section{Results}

\begin{figure*}

  \centering

  \begin{tabular}{cc}


    \includegraphics[width=70mm]{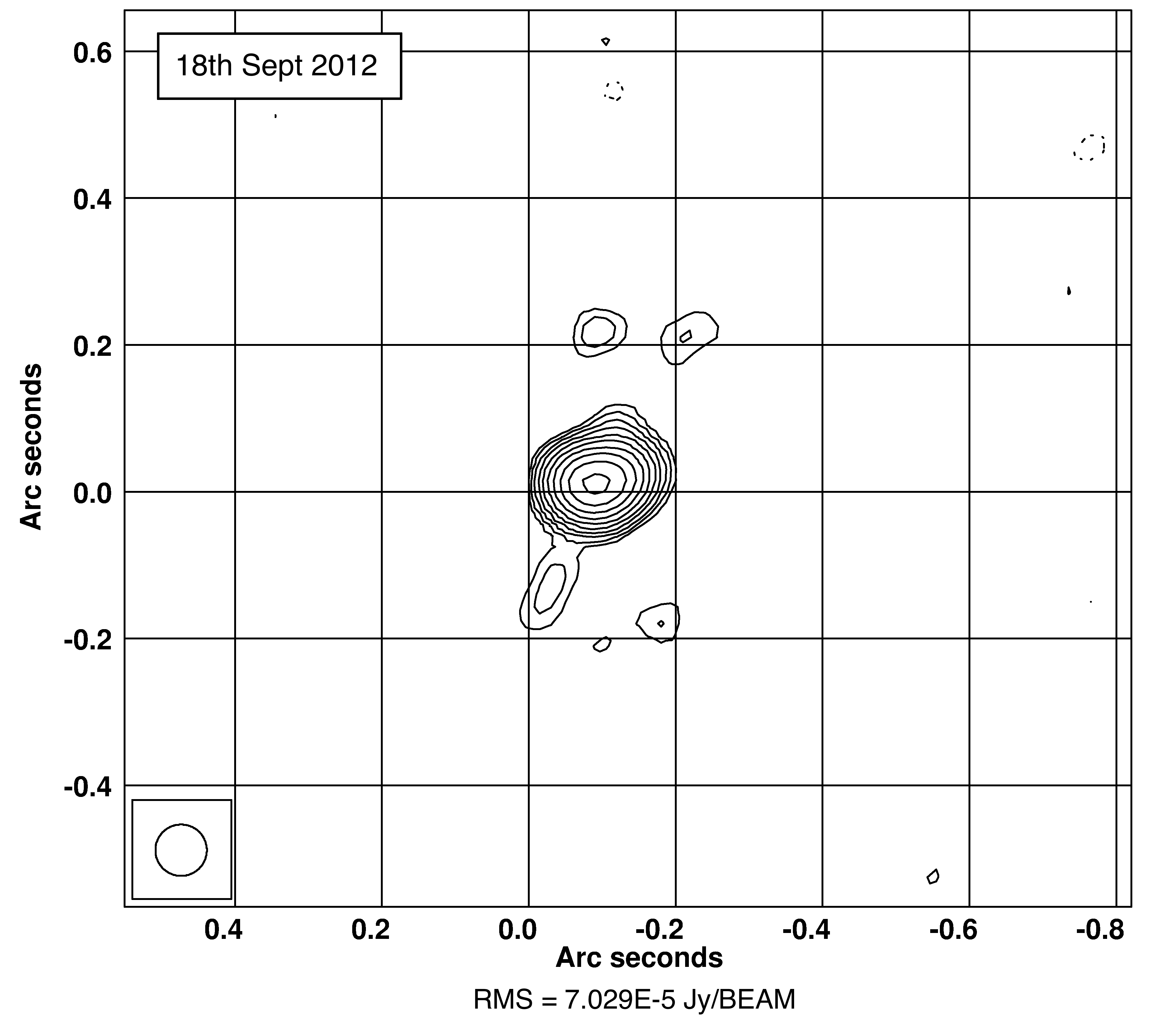}&

    \includegraphics[width=70mm]{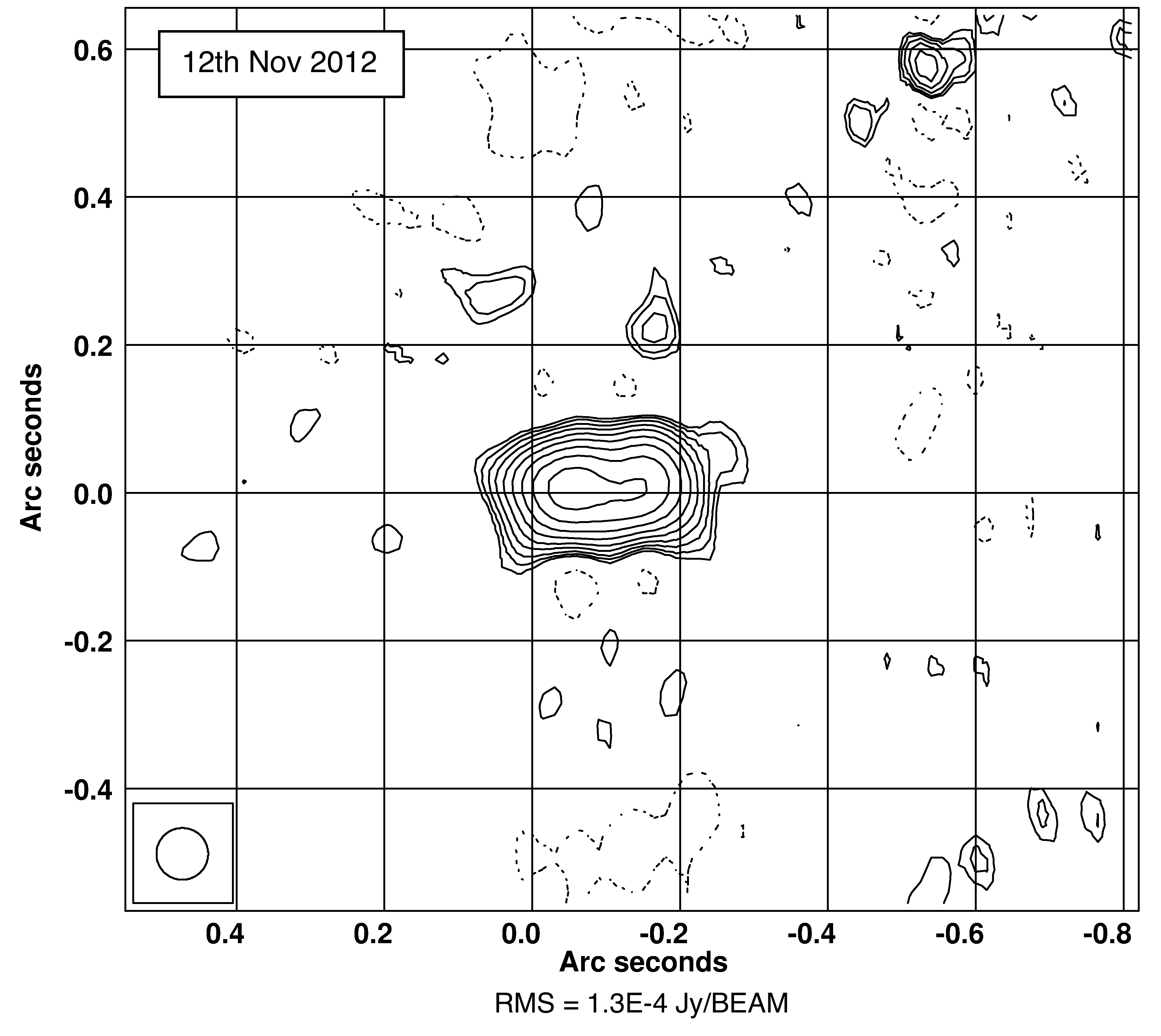}\\

    \includegraphics[width=70mm]{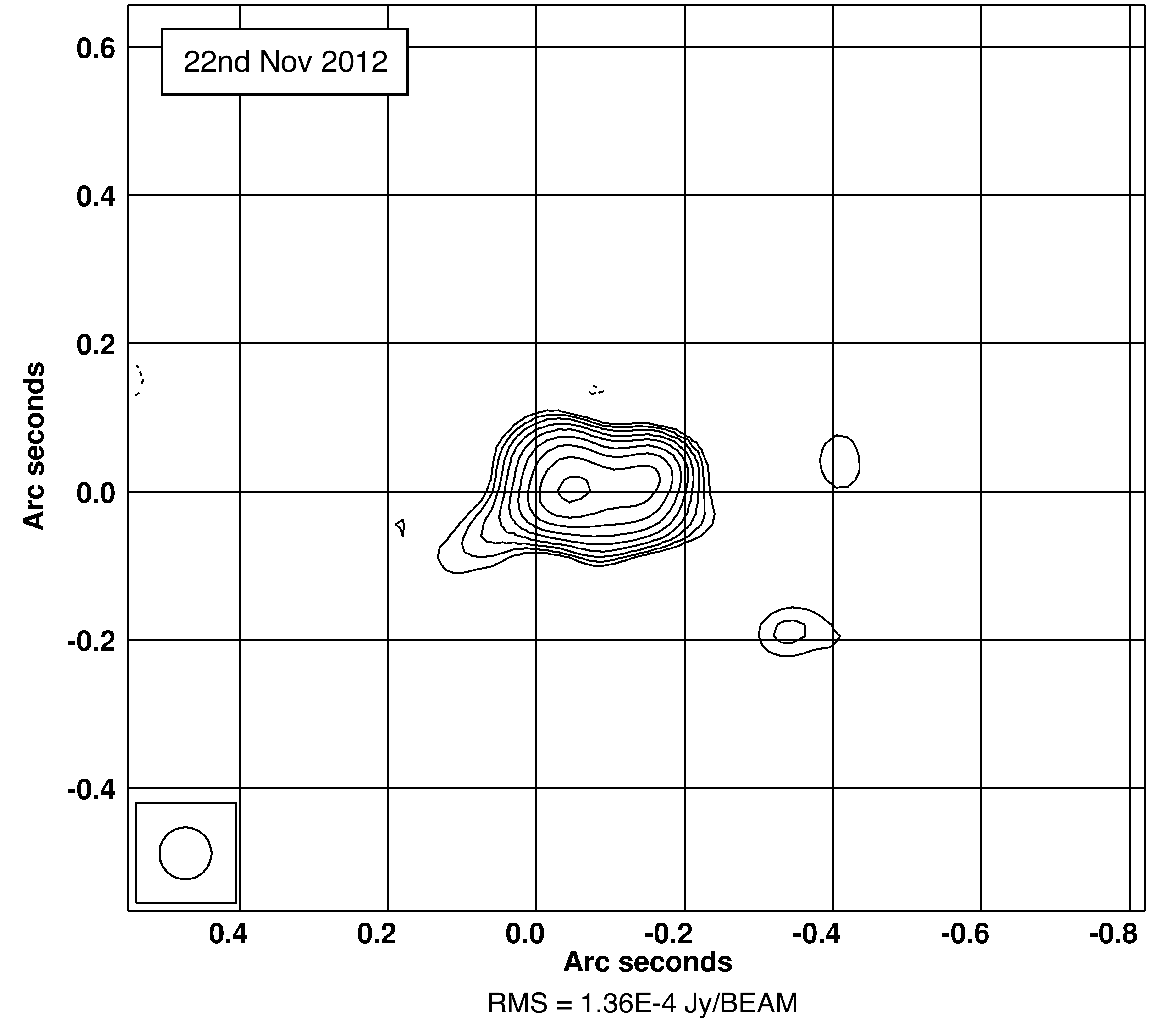}&

    \includegraphics[width=70mm]{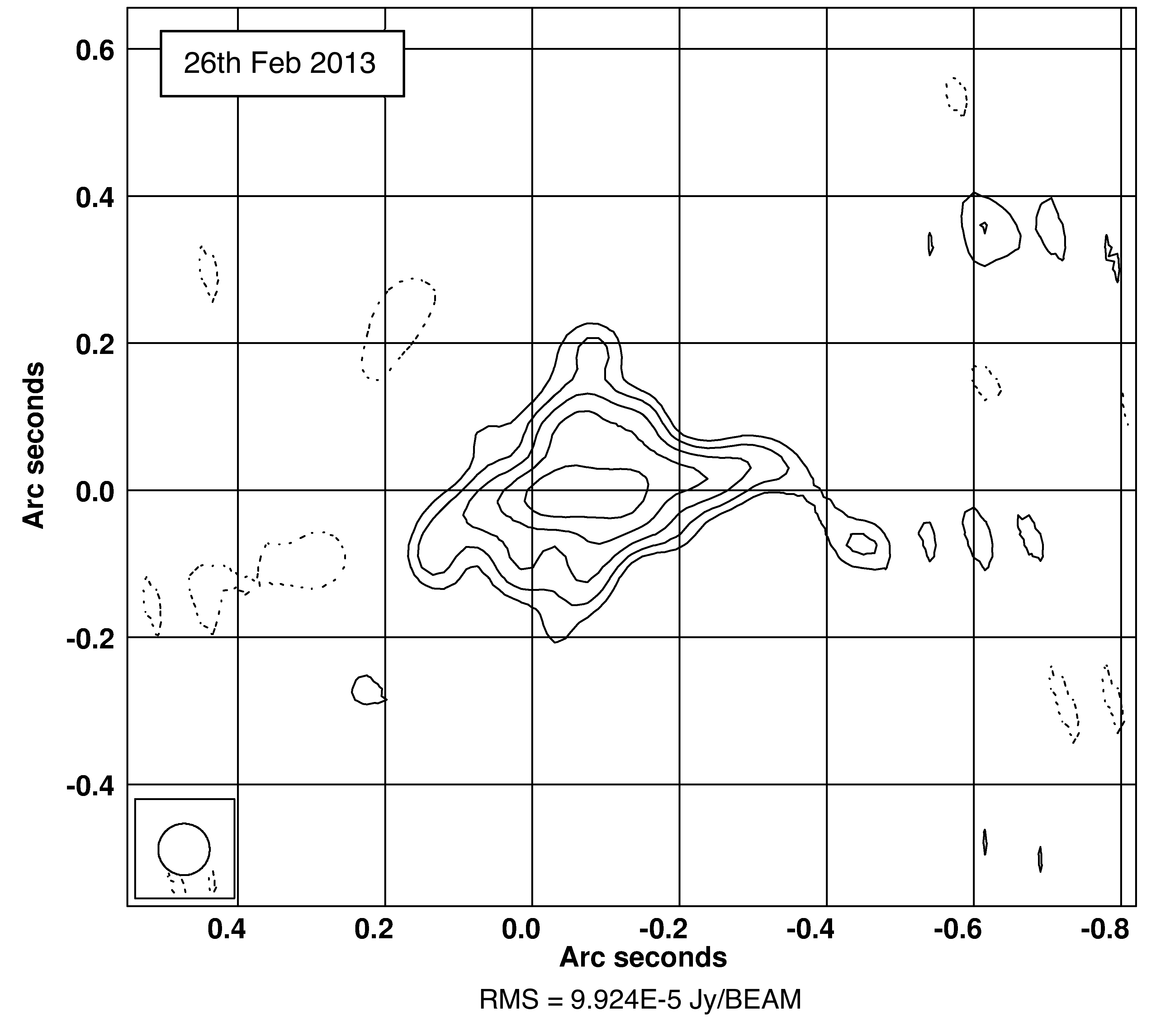}\\

    \includegraphics[width=70mm]{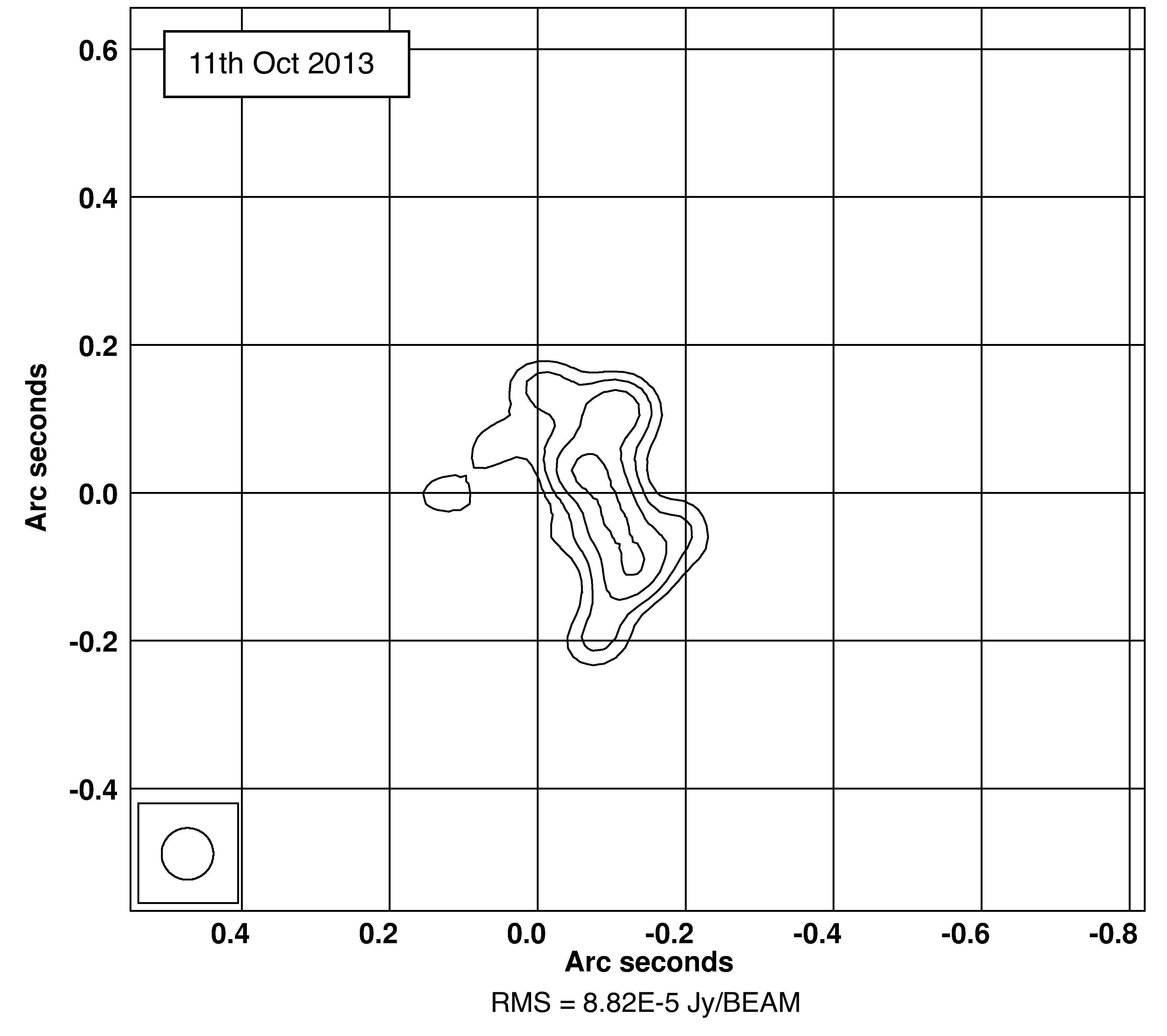}&

    \includegraphics[width=70mm]{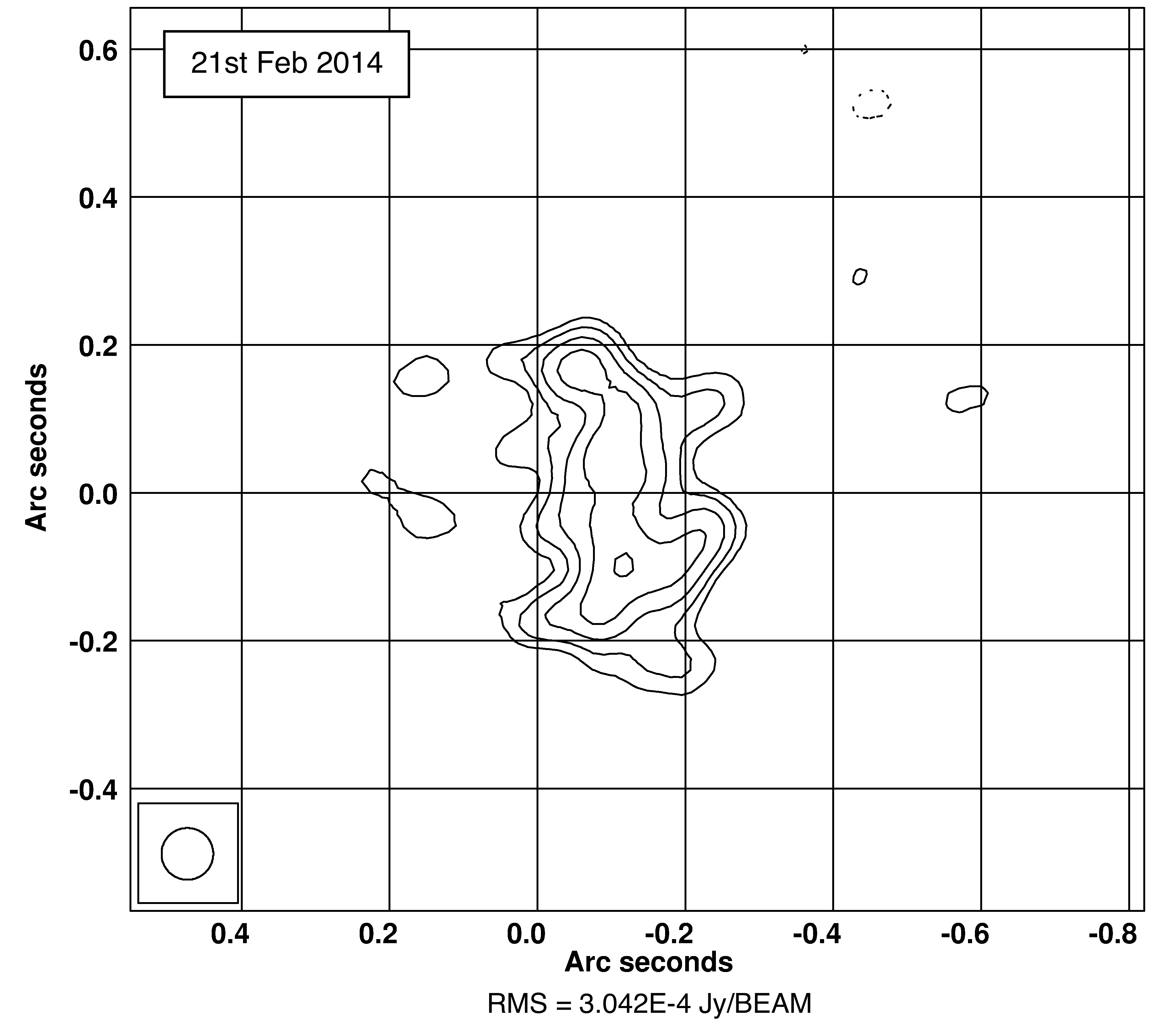}\\

  \end{tabular}
  \caption{\textit{Intensity contour plots of V959 Mon;
  from top-left: September 2012, 12th November 2012, 22nd
  November 2012, February 2013, October 2014, February 2014.
  The morphology can be seen to change from an east-west
  orientation, in the first four epochs, to what seems
  to be a north-south orientation, in the last two epochs.
  Each image was constructed using a 70 x 70 mas beam.
  The contour levels are equal to (-3, 3, 4.240, 6, 8.485,
  12, 16.97, 24, 33.94, 48, 67.88, 96) times the rms noise
  level (according to AIPS task IMSTAT) in the image.
  The origin of the images (0,0 arcsec) translates to a
  position with right ascension =
  06$^\circ$39$^\prime$38.606$^{\prime \prime}$ and
  declination = 05$^\circ$53$^\prime$52.830$^{\prime \prime}$. The RMS
  at each epoch is given below each image.}}
\label{figur}

\end{figure*}

Images of V959 Mon from each epoch are presented in
Figure \ref{figur}. Table \ref{Fluxes} shows the total
integrated flux for each epoch. Fluxes were calculated
using two different methods; using the AIPS tasks JMFIT
and IMSTAT. JMFIT fits an elliptical Gaussian to the source,
and integrates it to find the flux contained inside it.
IMSTAT calculates the total flux contained within the image. The average of the two results was taken,
and the uncertainty was taken to be the difference between
them. At C-Band, e-MERLIN has a theoretical sensitivity of
7 - 15 $\mu$Jy, assuming 12 hours on source and all antennas
in use. However, for some of the epochs presented here the source was observed for less than 12 hours (see Table \ref{times}), and the Lovell telescope was only in use for the February 2013 epoch. Furthermore, at every epoch some data was lost due to flagging of radio frequency interference (RFI), decreasing the sensitivity. The achieved sensitivities are shown in Figure \ref{figur}.

\begin{table}
 \centering
\begin{tabular}{| c | c | c |}
\hline
Epoch &  \begin{tabular}{@{}c@{}}Days past \\ Outburst\end{tabular} & \begin{tabular}{@{}c@{}} Total \\ Flux (mJy)\end{tabular} \\
\hline
18th Sept 2012 & 90 & 5 $\pm$ 1  \\
12th Nov 2012 & 150 & 19 $\pm$ 1  \\
22nd Nov 2012 & 159 & 27 $\pm$ 1 \\
26th Feb 2013 & 254 & 18 $\pm$ 1 \\
11th Oct 2013 & 482 & 8 $\pm$ 1 \\
21st Feb 2014 & 615 & 4 $\pm$ 1 \\
\hline
\end{tabular}
\caption{\textit{Total flux of V959 Mon at each epoch. The fluxes
were calculated with JMFIT and IMSTAT, and an average was taken.
The difference between the JMFIT and IMSTAT values was taken as the
uncertainty.}}
\label{Fluxes}
\end{table}

\begin{table}
 \centering
\begin{tabular}{| c | c | c |}
\hline
Epoch & Total Flux (Jy) & Error (Jy) \\
\hline
18th Sept 2012 & 0.316 & 0.001  \\
12th Nov 2012 & 0.295 & 0.001  \\
22nd Nov 2012 & 0.184 & 0.001 \\
26th Feb 2013 & 0.147 & 0.001 \\
11th Oct 2013 & 0.245 & 0.002 \\
21st Feb 2014 & 0.209 & 0.001 \\
\hline
\end{tabular}
\caption{\textit{Total integrated fluxes and associated
errors for the phase calibrator source 0645+0541 at each
epoch. These fluxes and errors were calculated using the
AIPS task JMFIT.}}
\label{phaseflux}
\end{table}

It was noted during the data reduction process that the
flux of the phase calibrator source (0645+0541) varied
from epoch to epoch, as outlined in Table \ref{phaseflux}.
The variations in flux were considered too great to be caused
by source variability (for example, its flux changed by ~0.11 Jy
in the 10-day period between 12th November 2012 and 22nd November 2012),
meaning the variation was probably caused by an instrumentation or
calibration issue; for example the hour angle of the flux calibrator source was different at every epoch, which may have caused the flux calibrations to vary from epoch to epoch. The flux of 0645+0541 was reported by
\citet{Chomiuk1} to be $\sim$ 0.175 Jy,
As such, all phase calibrator fluxes were corrected to
the flux of the observations made on 22nd November 2012, which
was closest to 0.175 Jy. The fluxes of V959 Mon were then
multiplied by the corrective factor used to normalize
the phase calibrator fluxes. Table \ref{Fluxes} shows
the corrected values.

\begin{figure*}
\centering
\includegraphics[width=0.7\textwidth]{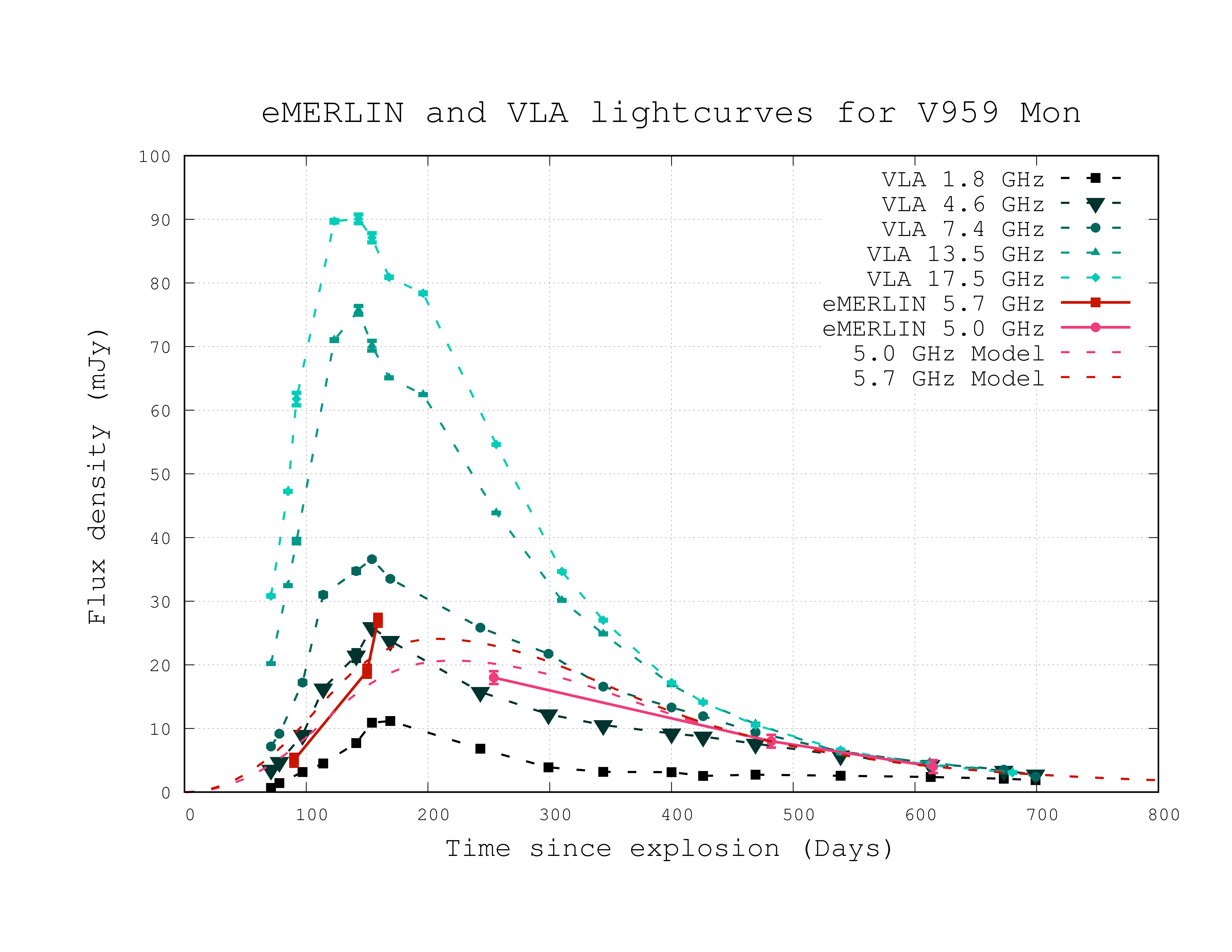}
\caption{\textit{e-MERLIN 5.7 and 5.0 GHz light curves for
V959 Mon, shown alongside multi-frequency VLA light curves
\citep{Chomiuk1}. All times are measured with reference to the
Fermi discovery of V959 Mon on June 19th 2012. For five of the six eMERLIN epochs, the observation spanned multiple days, so when plotting the light curve the median date was used.}}
\label{Curve}
\end{figure*}

The data in Table \ref{Fluxes} were used to construct
light curves for V959 Mon, shown in Figure \ref{Curve}.
While all the observations were made in the C-band,
two different central frequencies were used; 5.7 GHz
for the 2012 observations and 5.0 GHz for the post-2012
observations. As such, two separate light curves were
plotted. These light curves are in broad agreement with
the VLA light curve constructed by Chomiuk et al. (2014),
as shown in Figure \ref{Curve}.

Also shown in Figure \ref{Curve} is a comparison between the
light curve of V959 Mon as observed by
e-MERLIN, and a simulated light curve fit to the
data using the Hubble flow model. Using the best
fit parameters to the VLA light curves as reported
by Chomiuk et al. (2014), the model fitting program
was given initial guesses of $4 \times 10^{-5} M_{\odot}$
for the ejecta mass, $2300$ km s$^{-1}$ for the outer
velocity of the shell and $200$ km s$^{-1}$ for the
inner velocity of the shell (recall that the Hubble
flow model assumes an expanding spherical shell of ejecta).
The distance to the nova (1.5 kpc, \citealp{Munari}) and
the electron temperature of the ejecta ($2 \times 10^4$
K, \citealp{Chomiuk1}) were fixed. The fit returned best fit parameters of $1.1 \times 10^{-4} M_{\odot}$ for the ejecta mass, 2280 km s$^{-1}$  for the velocity of the outer edge of the shell and 490 km s$^{-1}$  for the inner velocity.

In the first four epochs (see Figure \ref{figur}), the
ejecta appear to be elongated in the east-west direction.
The elongation appears to become more pronounced with each
successive epoch.

In the subsequent epochs (October 2013 and February 2014),
the emission is elongated north-south rather than east-west.
It is possible that this north-south structure can be seen
to be emerging in the February 2013 epoch, which, though
predominately elongated from east to west, also appears
to have a north-south component (see Figure \ref{figur}).

\begin{figure}
\centering
\includegraphics[width=0.5\textwidth]{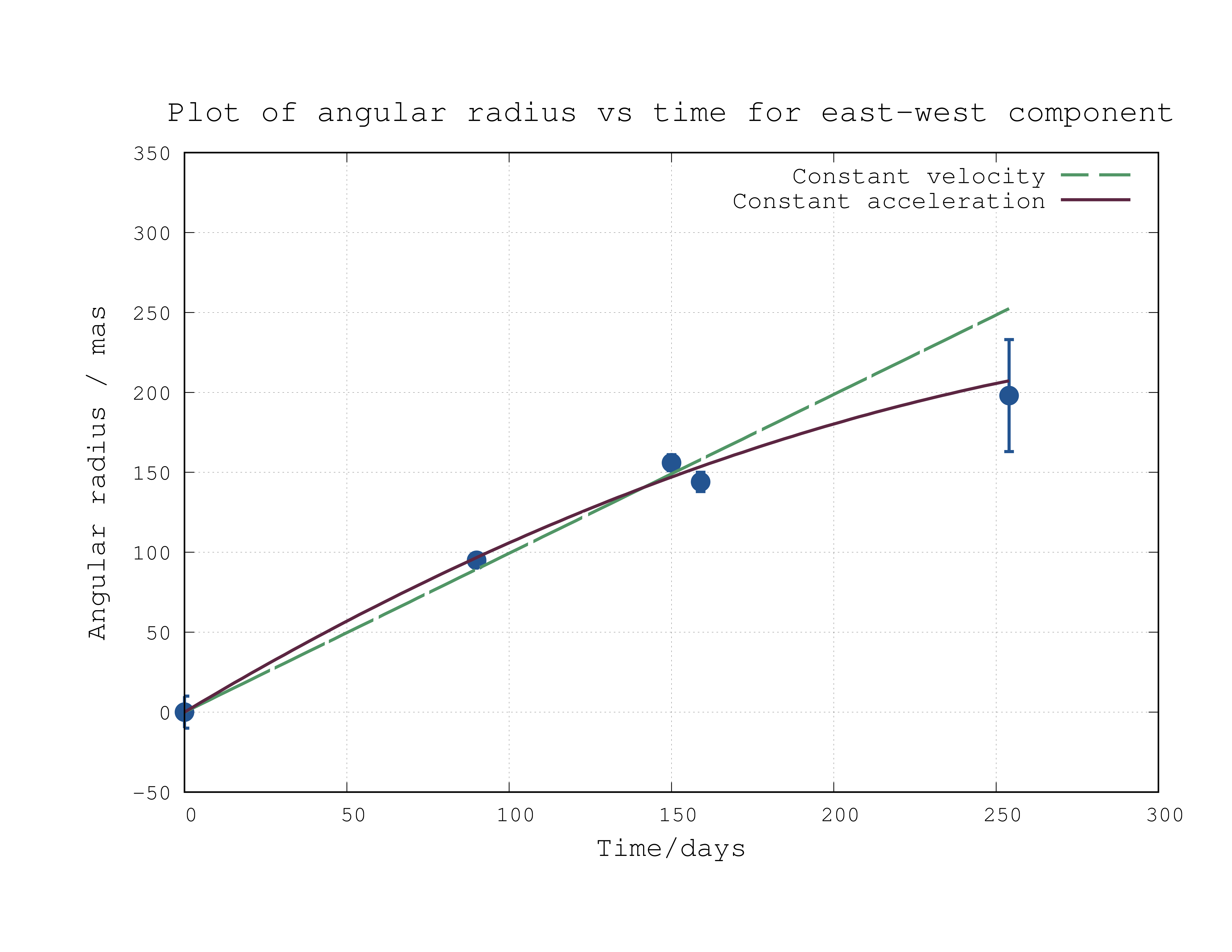}
\caption{\textit{Plots of angular radius (measured in the east-west direction)
vs time for the east-west phase of V959 Mon's expansion (using the September 2012, 12th November 2012, 22nd November 2012 and February 2013 epochs).}}
\label{EW}
\end{figure}

\begin{figure}
\centering
\includegraphics[width=0.5\textwidth]{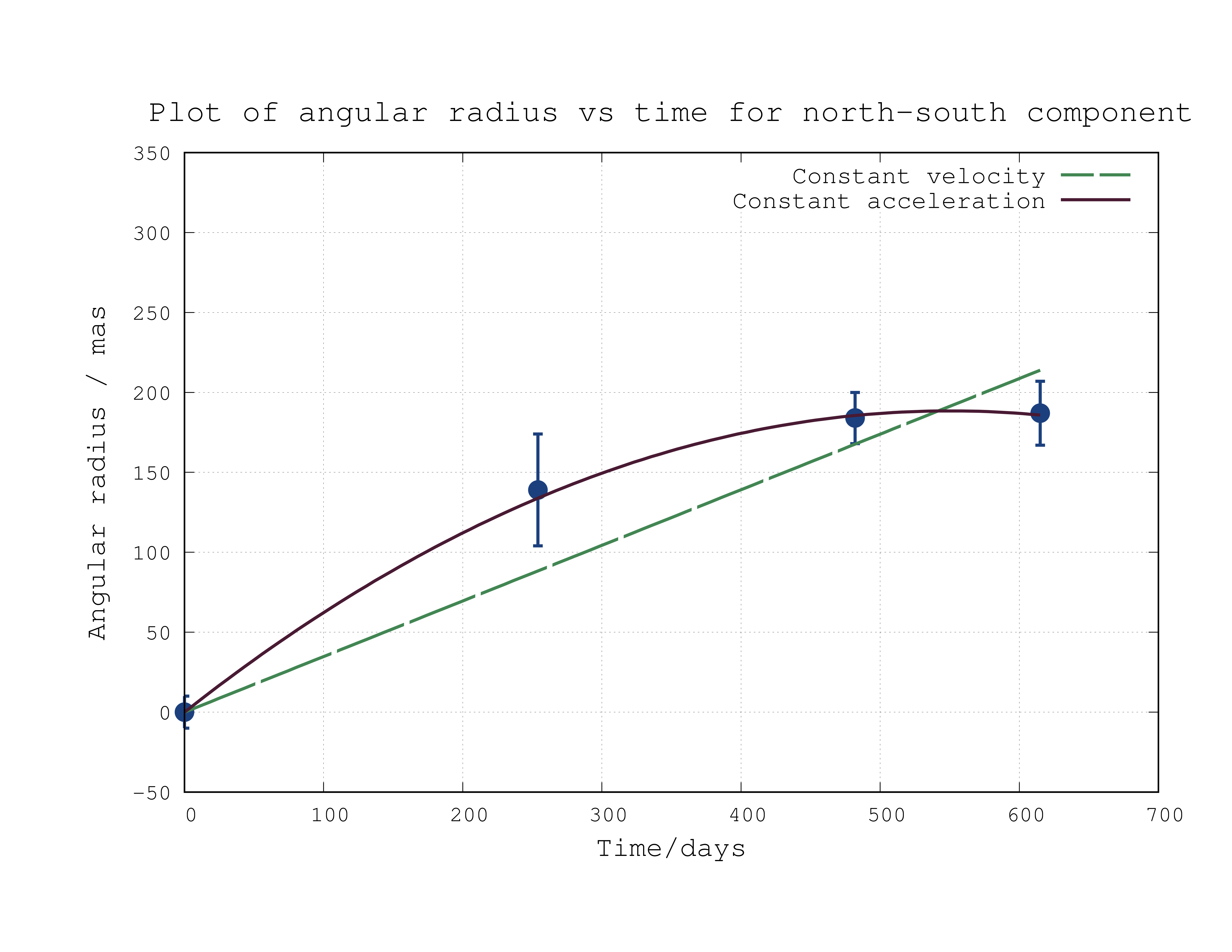}
\caption{\textit{Plots of angular radius (measured in the north-south
direction) vs time for the north-south phase of V959 Mon's expansion (using the February 2013, October 2013 and February 2014 epochs).}}
\label{NS}
\end{figure}

In order to further our understanding of V959 Mon's
expanding morphology, plots were made of radius
against time for both the east-west component and
the north south component. The angular radius of the
ejecta at each epoch was measured from the images shown in
Figure \ref{figur}, using the 4-$\sigma$ contours, and using
the difference between the 3- and 5-$\sigma$ contours
as the $\pm 1 \sigma$ uncertainty (since 3-$\sigma$ was considered to be
the lowest believable contour, it was taken as the
lower limit for the uncertainty in the radius measurement).
For each image, the radii were
measured along the line containing the peak flux. For
the first three epochs, the radii were measured from east
to west (where the ejecta was most extended), and for the
final two, the radii were measured from north to south. For
the fourth epoch (February 2013), the radius was measured
both in the north-south direction and in the east-west
direction, as evidence of both components could be seen
in the image. In both cases, a radius of zero was assumed at the
beginning of the outburst.
The measured radii are shown in Tables \ref{EW_Rad}
and \ref{NS_Rad}.

The plots of radius vs time are shown in Figures
\ref{EW} and \ref{NS}. For both the east-west and
north-south phase, two lines were fit to the data;
$r(t) = vt$, assuming a constant velocity equal to
$v$, and $r(t) = ut + \frac{1}{2}at^2$, assuming a
constant acceleration equal to $a$, with an initial
velocity of $u$. For the east-west component of ejecta,
the constant-velocity fit returned a value of
$v = 2500 \pm 110$ km s$^{-1}$ for the velocity,
and the constant-acceleration fit returned values
of $a = - 8 \pm 5$ km s$^{-2}$ and $u = 3160 \pm 350$
km s$^{-1}$ for the acceleration and initial velocity
of the ejecta. Since the uncertainty on the acceleration was so high, we conclude that
the constant velocity
fit provides a more realistic description of the
dynamics of the east-west component.

\begin{table}
 \centering
\begin{tabular}{| c | c | c |}
\hline
Epoch & \begin{tabular}{@{}c@{}} Days Since \\ Outburst \end{tabular} & \begin{tabular}{@{}c@{}}Angular \\ Radius\\ (mas)\end{tabular}  \\
\hline
18th Sept 2012 & 91 & 95 $\pm$ 4  \\
12th Nov 2012 & 146 & 156 $\pm$ 5  \\
22nd Nov 2012 & 156 & 144 $\pm$ 6 \\
26th Feb 2013 & 252 & 198 $\pm$ 35 \\
\hline
\end{tabular}
\caption{\textit{Angular radii (in milliarcseconds)
of the ejecta at each epoch during the east-west phase
of the expansion (September 2012, November 12th 2012,
November 22nd 2012, and February 2013.)}}
\label{EW_Rad}
\end{table}

\begin{table}
 \centering
\begin{tabular}{| c | c | c |}
\hline
Epoch & \begin{tabular}{@{}c@{}} Days Since \\ Outburst \end{tabular} & \begin{tabular}{@{}c@{}}Angular \\ Radius\\ (km/s)\end{tabular}  \\
\hline
26th Feb 2013 & 252 & 139 $\pm$ 35 \\
11th Oct 2013 & 479 & 184 $\pm$ 16 \\
21st Feb 2014 & 612 & 187 $\pm$ 20 \\
\hline
\end{tabular}
\caption{\textit{Angular radii (in milliarcseconds) of
the ejecta at each epoch during the north-south phase
of the expansion (February 2013, October 2013, and February 2014).}}
\label{NS_Rad}
\end{table}

For the north-south component, a value of $v = 900 \pm 90$
km s$^{-1}$ was returned by the constant-velocity fit,
and values of $a = -3 \pm 0.2$  km s$^{-2}$ (indicating a
deceleration) and $u = 1780 \pm 50$ km s$^{-1}$ were
returned by the constant-acceleration fit. In this
instance, assuming a constant acceleration appeared to be a better model (see Figure \ref{NS}).
Comparing the average expansion velocities of the east-west and north-south components (i.e.
the constant velocity fits to the radius vs time plots), we find that on average the east-west
component had a significantly faster expansion (by about 1700 km s$^{-1}$).

\section{Discussion}

Using images of the ejecta morphology observed in V959 Mon by e-MERLIN, we have attempted to learn more about the behaviour of this complex object. Our method of estimating
the expansion velocity differs from that of \citet{Linford},
who made their estimation by fitting SHAPE models to
the observed ejecta. However, our estimation of the velocity of
the north-south component lies within the error bars of
\citet{Linford}, who estimate that the maximum velocity of
the north-south component is 1200 km/s.

We note here, however, that if the
ejecta were expanding according to the Hubble flow model,
an apparent decrease in expansion velocity from epoch to epoch may also be
measured. This is because (as noted in \citealp{Linford}) as
the shell expands outwards, it becomes optically thin, first
at the outer edges and
gradually all the way through. As a result, the effective
photosphere, which had been propagating outwards from the
WD, would move back towards it. If expansion velocities
were being measured from the size of the observed ejecta,
a decrease in velocity could be observed at this time. It is possible that this might contribute to our observation of a decelerating outflow in the N-S direction.

MERLIN observations by \citet{Heywood1} of V723 Cas showed similar developments in the ejecta morphology to what we have observed here; a seeming north-south elongation which then gave way to an east-west structure. However, simulations of the effect of MERLIN UV coverage on a simulated spherically symmetric density distribution by \citet{Heywood2} showed spurious aspherical features which were strikingly similar to the MERLIN observations of V723 Cas. They concluded was that it is necessary to be very careful when analysing resolved MERLIN images of nova ejecta, as seemingly complex aspherical ejecta could simply be the result of instrumental effects. V959 Mon is close to equatorial with a declination of around 6.5$^\circ$ and imaging can be affected by the e-w bias of the array resulting in the beam having significant n-s structure and enhanced sidelobes.   

The improved UV coverage of e-MERLIN with its much higher bandwidth with respect to MERLIN may have helped mitigate these effects, nonetheless when analysing V959 Mon's morphology, it was necessary to explore the possibility that the variable bipolar distortions we observed in V959 Mon were spurious features resulting from the e-MERLIN UV coverage.  Several steps were taken to determine whether or not this was the case. First, a spherically symmetric density distribution was calculated using the best fit parameters from the Hubble flow fit to the e-MERLIN lightcurve of V959 Mon (as described in the Results section), for each of the epochs at which V959 Mon was observed. The ejecta were given a mass of $1.1 \times 10^{-4} M_{\odot}$, as predicted by the model fit. The outer and inner radii of the spherical shell were determined at each epoch by multiplying the velocities returned by the model fit to the light curve (2280 and 490 km s$^{-1}$ respectively) by the amount of time that had elapsed between the Fermi discovery and the epoch in question. The ejecta were given a density distribution that varied with the square of the radial difference from the white dwarf surface, via (see \citealp{Hjellming1990}): $\rho(r) = M/4 \pi (R_o - R_i) r^2$, where \textit{M} is the mass of the spherical shell, ${R_o}$ is its outer radius, ${R_i}$ is its inner radius and \textit{r} is the radial distance from the WD surface.

Radio emission was simulated from these density distributions by integrating the equations of radiative transfer along the line of sight through the ejecta (see \citealp{Heywood2}). The simulated emission was then sampled by a reconstruction of the e-MERLIN UV coverage on the dates of the observations presented in this paper (using a method adapted from \citealp{ALMA}), and images of the simulated ejecta were made.

\begin{figure*}

  \centering

  \begin{tabular}{ccc}


    \includegraphics[width=40mm]{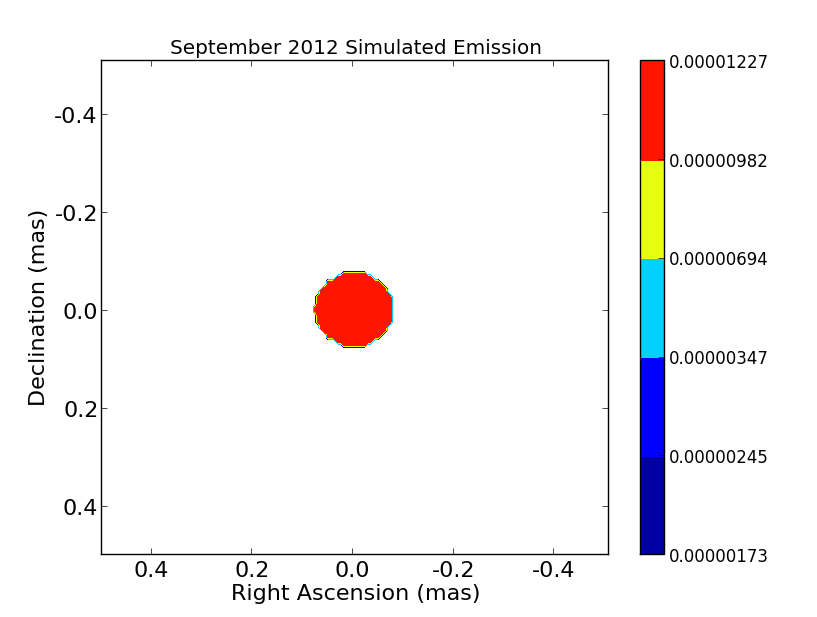}&

    \includegraphics[width=40mm]{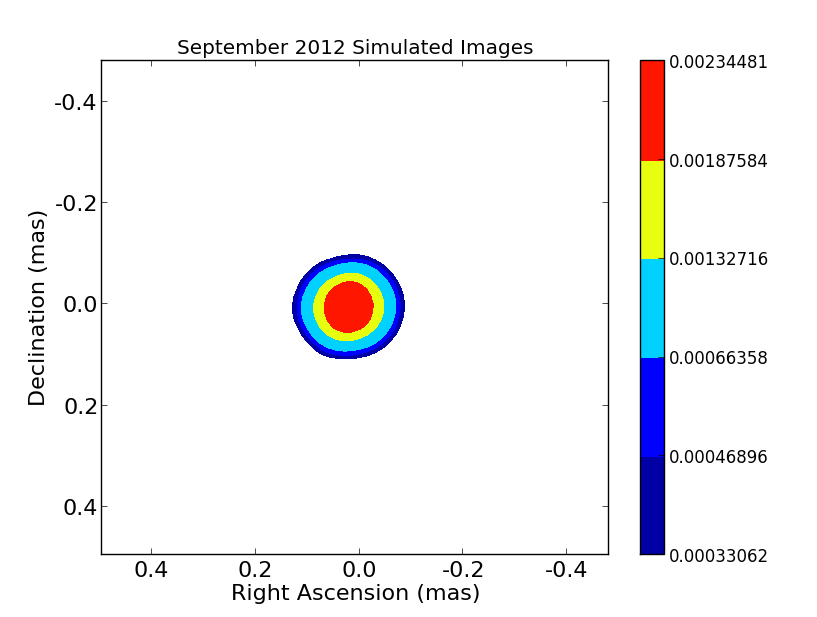}&

    \includegraphics[width=40mm]{aipsplot0.png}\\

    \includegraphics[width=40mm]{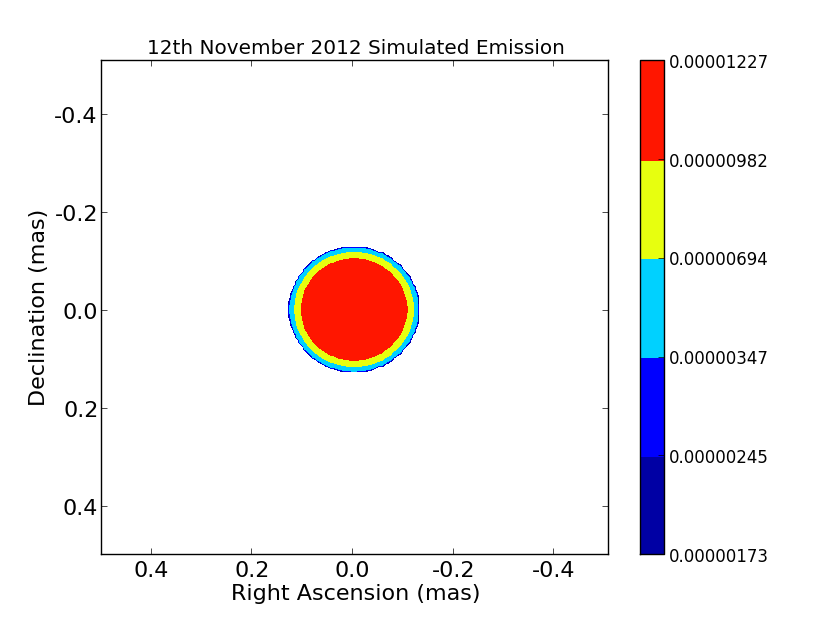}&

    \includegraphics[width=40mm]{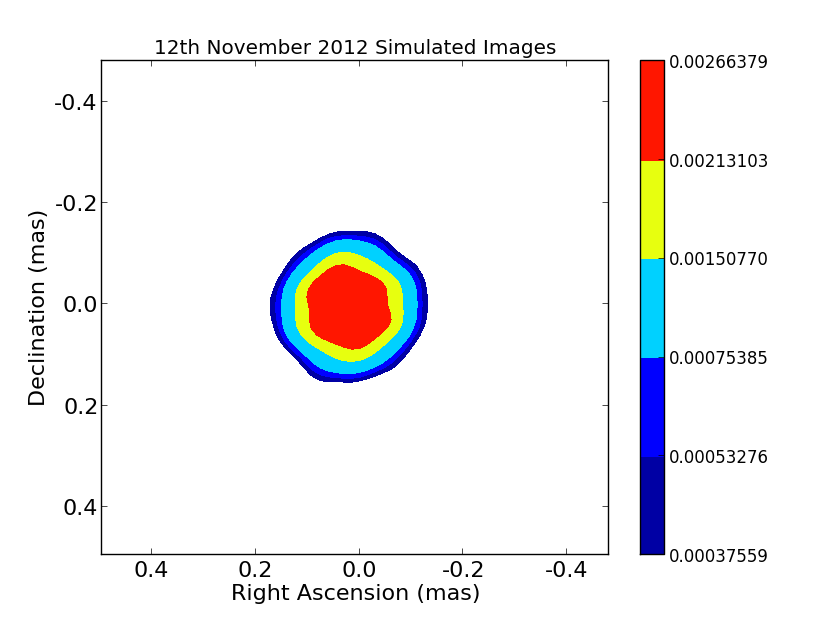}&

    \includegraphics[width=40mm]{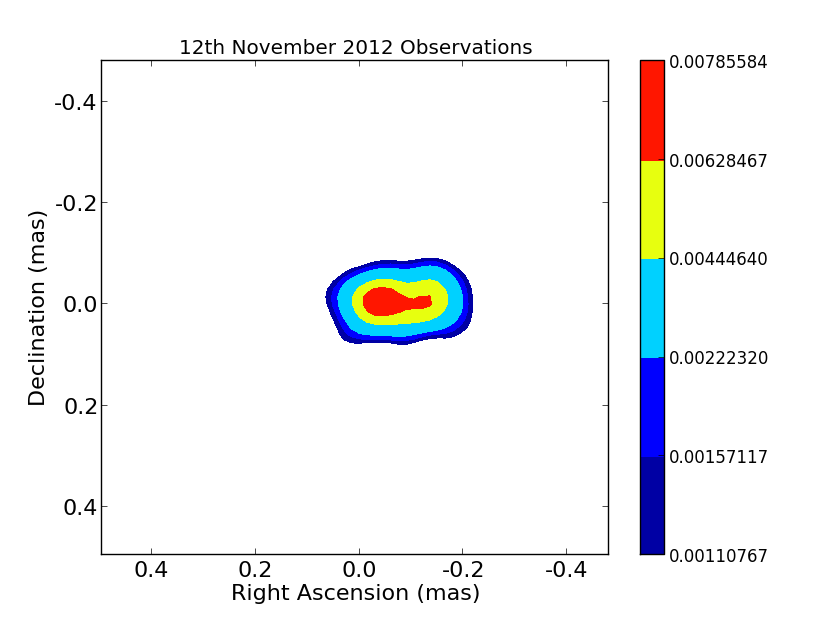}\\

    \includegraphics[width=40mm]{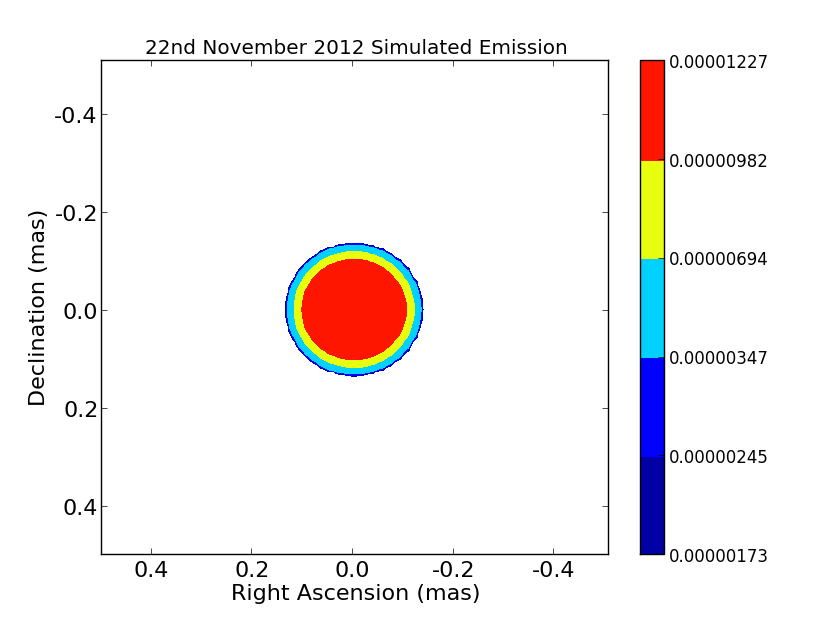}&

    \includegraphics[width=40mm]{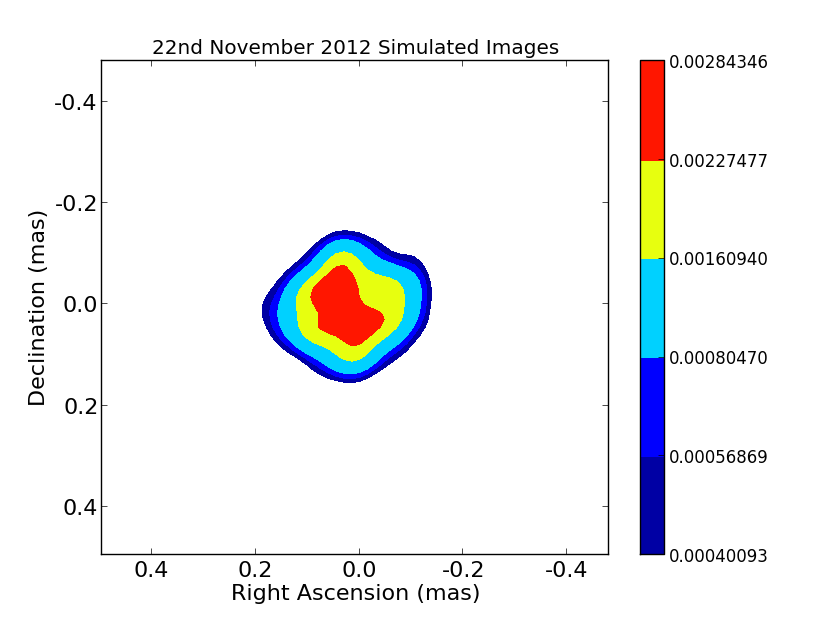}&

    \includegraphics[width=40mm]{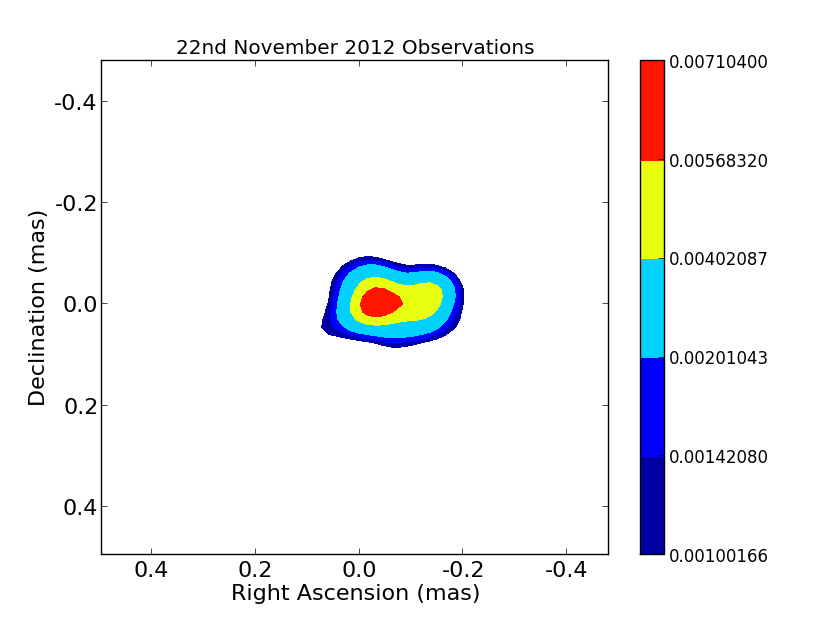}\\

    \includegraphics[width=40mm]{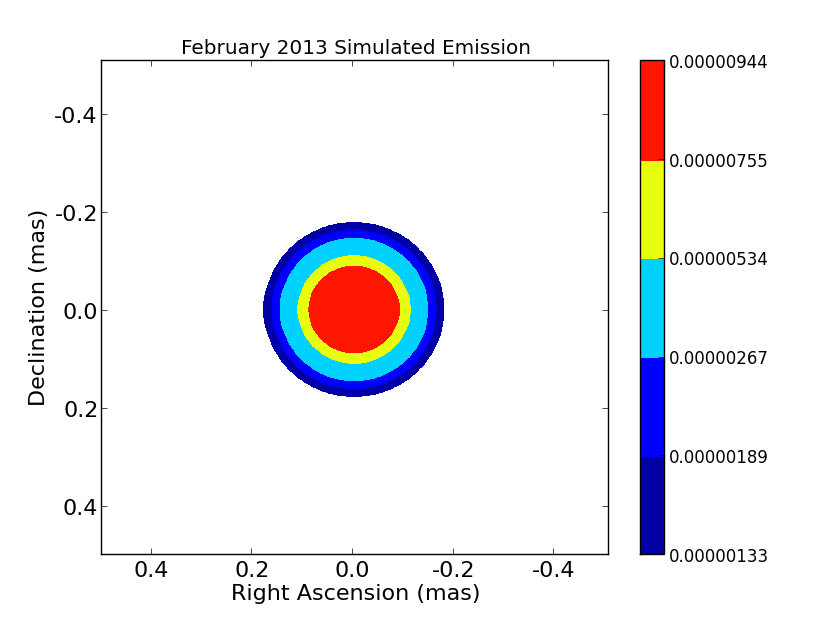}&

    \includegraphics[width=40mm]{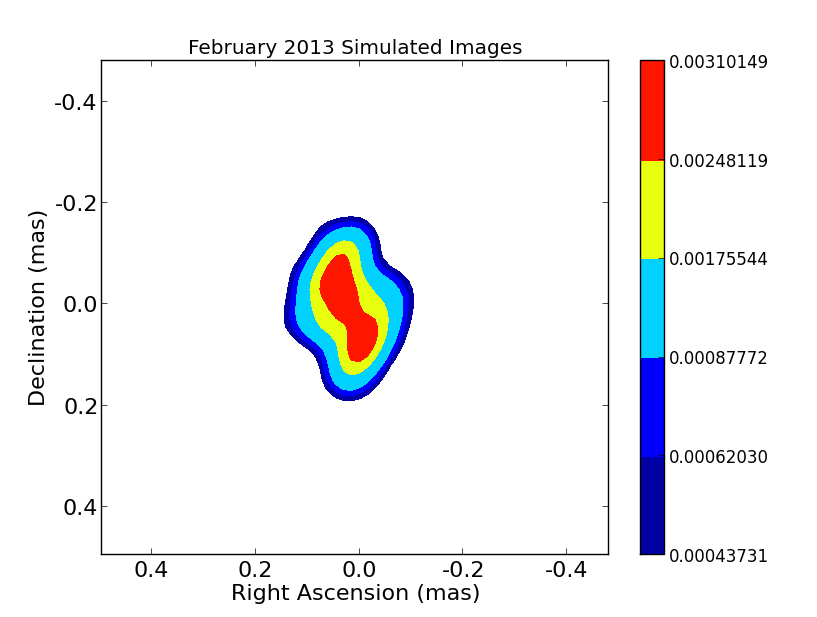}&

    \includegraphics[width=40mm]{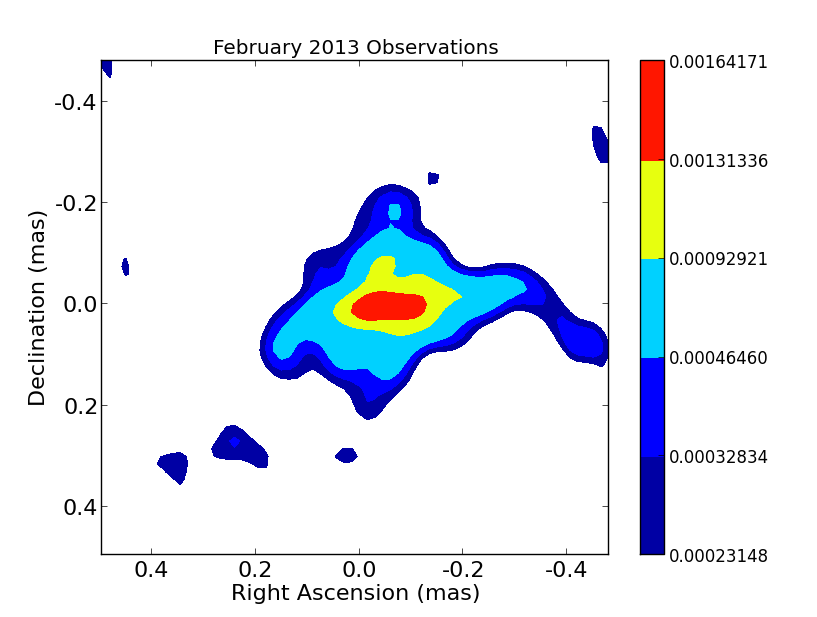}\\

    \includegraphics[width=40mm]{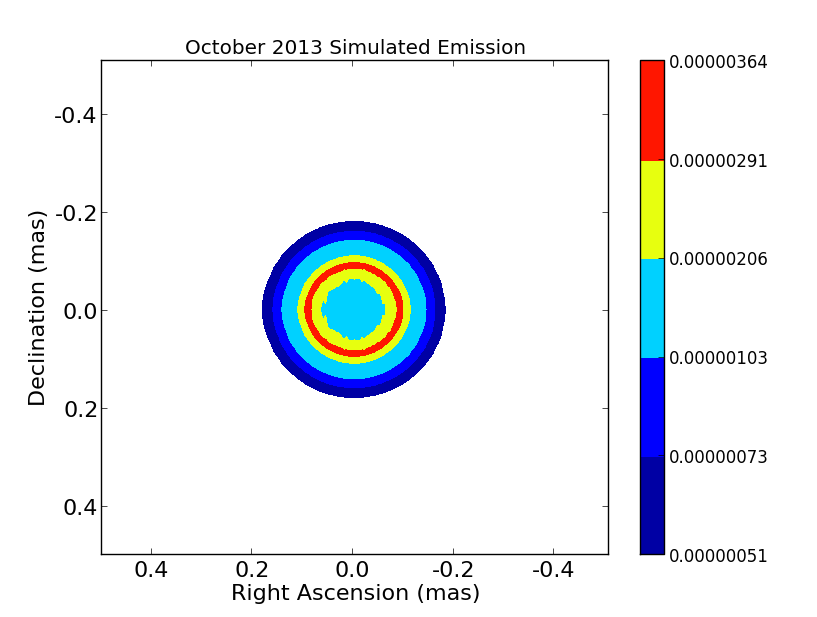}&

    \includegraphics[width=40mm]{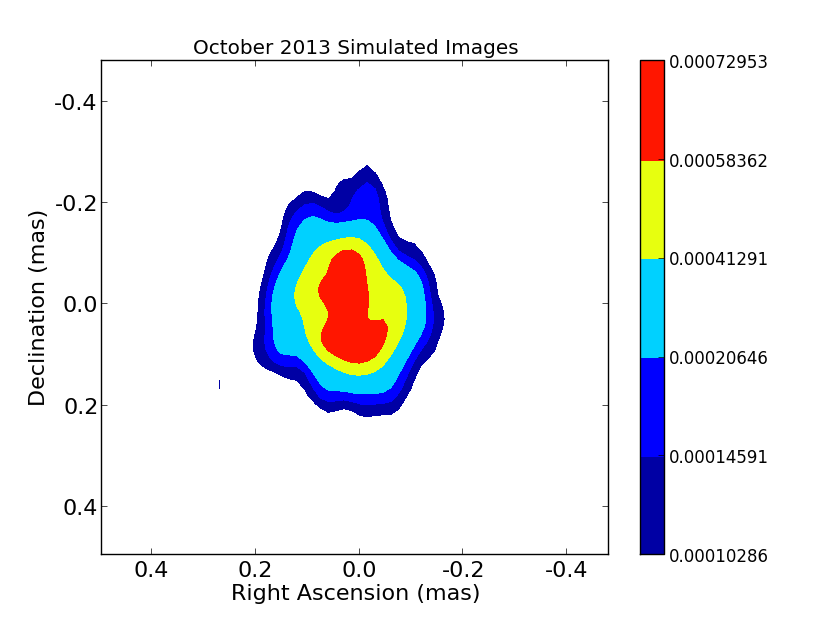}&

    \includegraphics[width=40mm]{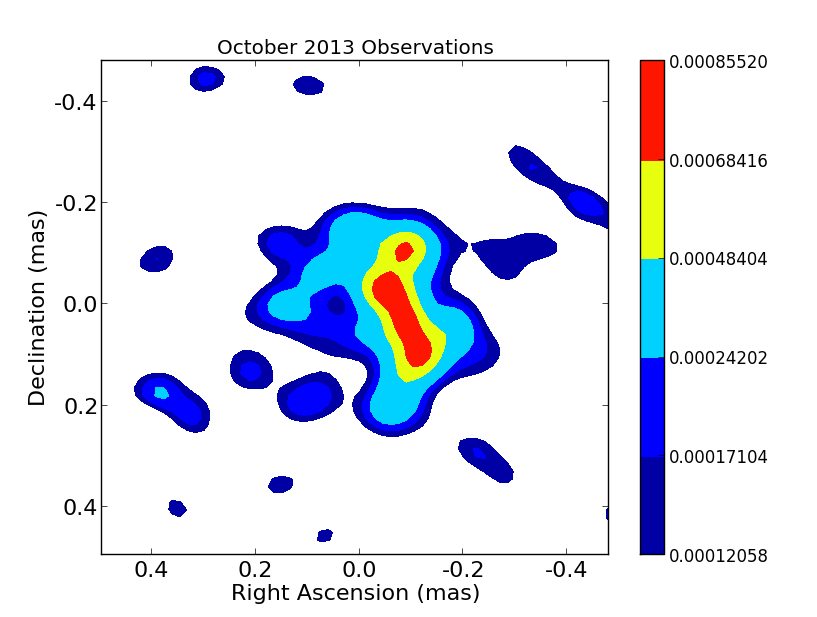}\\

    \includegraphics[width=40mm]{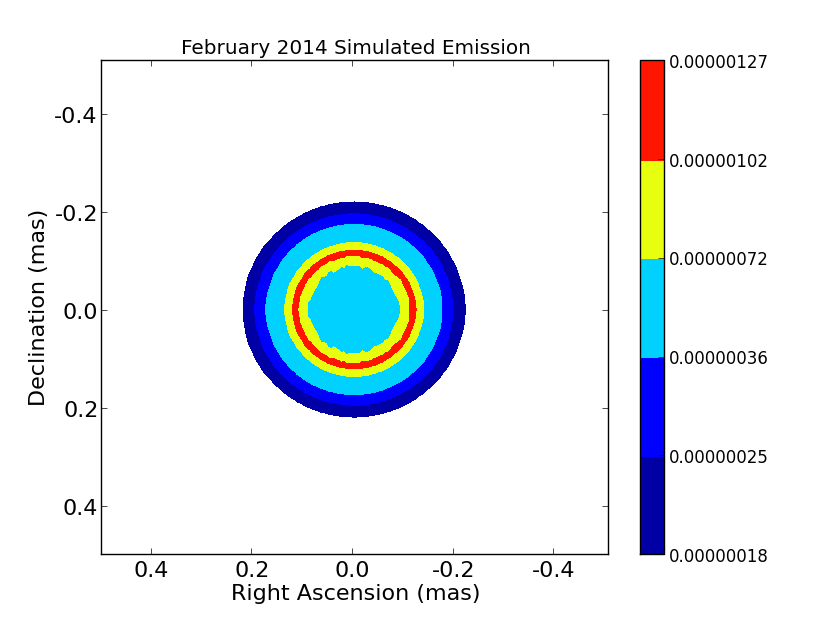}&

    \includegraphics[width=40mm]{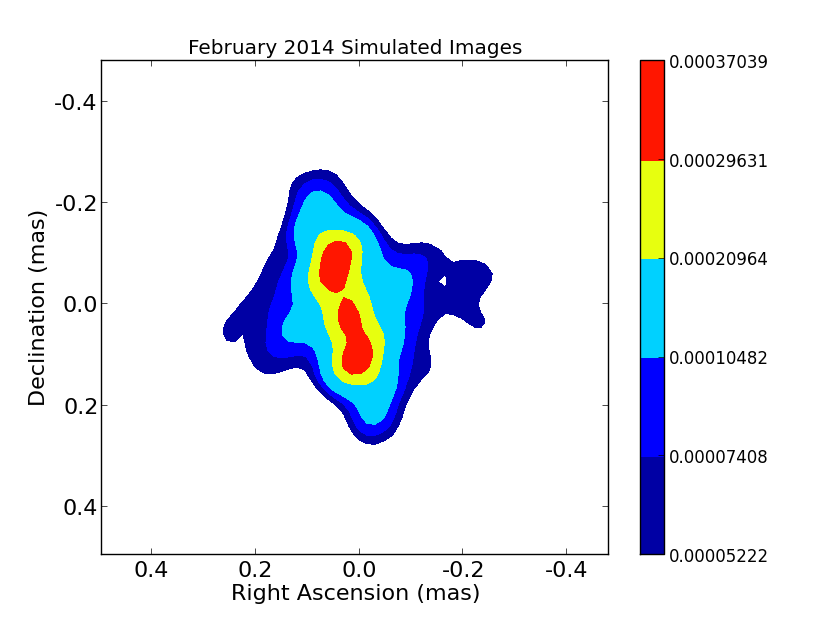}&

    \includegraphics[width=40mm]{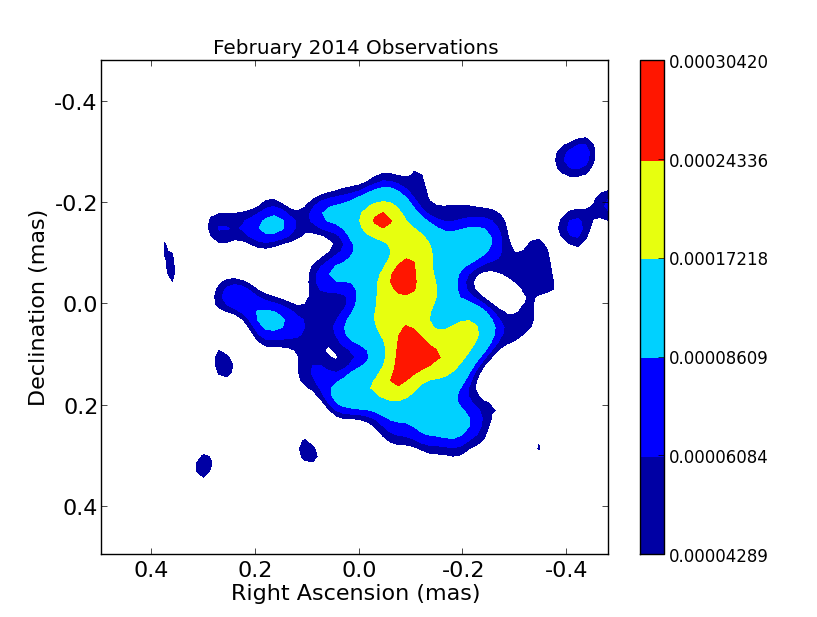}\\

  \end{tabular}
  \caption{\textit{Left: Simulated emission from an expanding spherical shell of ejecta corresponding to each epoch of observations presented here, using parameters predicted by a Hubble flow fit to the e-MERLIN light curve. Middle: simulated e-MERLIN images of the emission on the left, constructed by sampling the emission on the left with reconstructions of the e-MERLIN UV coverage on the dates that the observations presented in this paper were made. Right: e-MERLIN observations of V959 Mon (as presented in Figure \ref{figur}.) The images in the middle and on the right were constructed with a cellsize of 0.015 arcsec/pixel, using a 0.07x0.07 arcsec beam, and are centred on the peak of the flux. The simulated emission on the left was constructed with a cellsize of 0.0051692 arcsec/pixel. The colourbar to the right of each plot indicates the flux, which is in Jy/Beam. The contour levels in each case were set to 0.141,0.2,0.283,0.566,0.8,1 times the peak flux. }}
\label{simobs}

\end{figure*}

Figure \ref{simobs} shows the simulated spherical radio emission (on the left) and the e-MERLIN images of the simulated emission (in the middle), alongside the corresponding e-MERLIN observations of V959 Mon (on the right). For the first four epochs (September 2012, 12th November 2012 and 22nd November 2012 and February 2013), extension from east to west can be seen in the observations of V959 Mon that is not present in the images of the simulated spherical emission, suggesting that the observed east-west extension is real and not just a consequence of limited UV coverage. However, the simulated image of the spherical shell (by this time appearing as a ring in the image) at the February 2013 epoch has a pronounced north-south distortion. This means that we can be less confident that the north-south feature observed in February 2013 is a true representation of the source structure. This distortion continues in the October 2013 and February 2014 simulations.

It should of course be noted that \citet{Chomiuk1} and \citet{Linford} also detected a change from east-west to north-south in the orientation of V959 Mon's ejecta, using multi-frequency VLA observations spanning 4.5 to 36.5 GHz. This would support interpreting the observed north-south extension in the e-MERLIN imaging as real. Furthermore, the extent of the n-s emission in the simulated observations (middle column of Figure \ref{simobs}) can be seen to be related to the diameter of the ring in the model images (left column) suggesting that measurements made from the observed images can still provide information on the size of the source even in a case where the detailed structure might be difficult to determine.  

\section{Conclusion}

High resolution e-MERLIN observations have been
presented here of V959 Mon (Nova Mon 2012) at six
epochs ranging from September 2012 to February 2014.
The observations were made at C-band ($\sim$5 GHz)
frequency. Resolved images of Nova Mon 2012 show a
complex, aspherical, evolving morphology, which
appears to change its orientation from an east-west elongation
to a north-south elongation. The e-MERLIN light
curve at C-band is in good agreement with VLA
observations of Nova Mon 2012 at higher and lower
frequencies.

The variable morphology of Nova Mon 2012 as observed with e-MERLIN and presented here is consistent
with the observations of \citet{Chomiuk1} and \citet{Linford}. The e-MERLIN
observations presented here also cover the period between
days 199 and 615 post-Fermi-discovery, during which the
ejecta changes from being elongated east-west to north-south.
The high resolution of
the e-MERLIN images presented here allowed us to
directly measure the angular radius of V959 Mon's
ejecta at each epoch, and estimate expansion velocities (and
accelerations) for the east-west and north-south components
of V959 Mon's ejecta. Our measurements of the expansion velocities of these two components indicate that they display
different dynamics, and hence are likely driven by different
mechanisms.

However, the change in orientation from east-west to north-south that we observed in V959 Mon's ejecta bore some similarity to MERLIN observations by \citet{Heywood1} of V723 Cas, in which a change in ejecta orientation was also  observed. A follow-up analysis by \citet{Heywood2} found that the complex structure seen in the ejecta of that source could be reproduced by applying MERLIN UV coverage to a spherically symmetric emission model, indicating that the unusual features seen in the observations were not real.

To investigate whether or not the complex morphology we have observed in V959 Mon can be believed, we have simulated spherical emission at times corresponding to the six epochs of e-MERLIN observations presented here. We have constructed images of this simulated emission by sampling it with reconstructions of the e-MERLIN UV coverage on the dates when the observations presented here were made. The simulated images do not have any significant east-west elongation but do feature a north-south distortion at the February 2013, October 2013 and February 2014 epochs. As such, we conclude that the observed east-west elongation is indicative of the actual source structure whilst we can be less certain of the north-south elongation. Although we note that \citet{Chomiuk1} and \citet{Linford} do also observe a north-south elongation in VLA imaging.

To further our understanding of V959 Mon's unusual behaviour, it will be of use to testthe model proposed by \citet{Chomiuk1} using radio emission models of aspherical ejecta. By constructing density
distributions representative of the ejecta behaviour proposed by \citet{Chomiuk1}, we can investigate whether or not such behaviour could result
in the radio emission we have observed, using plausible values for parameters such as ejecta mass and velocity. We will present the findings
of this study in a following paper.

\section{Acknowledgments}

e-MERLIN is a National Facility operated by the
University of Manchester at Jodrell Bank Observatory on
behalf of the UK Science and Technology Facilities Council
(STFC).

FH acknowledges the support of an STFC-funded studentship.

We thank the anonymous referee for their useful and insightful
comments.

\bsp

\label{lastpage}

\end{document}